\documentclass[aps,prd,
twocolumn,
nofootinbib,floatfix]{revtex4}

\usepackage{graphicx}
\usepackage{epsfig}
\usepackage{amssymb}

\usepackage{epstopdf} 

\newcommand{\be}{\begin{eqnarray}}
\newcommand{\ee}{\end{eqnarray}}

\newcommand\nn{\nonumber}

\newcommand{\mat}{\left ( \begin{array}{cc}}
\newcommand{\emat}{\end{array} \right )}
\newcommand{\vect}{\left ( \begin{array}{c}}
\newcommand{\evect}{\end{array} \right )}

\newcommand\hatmu{\hat{\mu}}
\newlength{\bredde}
\def\slash#1{\settowidth{\bredde}{$#1$}\ifmmode\,\raisebox{.15ex}{/}
\hspace*{-\bredde} #1\else$\,\raisebox{.15ex}{/}\hspace*{-\bredde} #1$\fi}
\def\Eq#1{Eq.~(\ref{#1})}

\begin{document}

\title{The Sign Problem via Imaginary Chemical Potential}
\author{K. Splittorff}
\affiliation{The Niels Bohr Institute, Blegdamsvej 17, DK-2100, Copenhagen {\O}, Denmark}
\author{ B. Svetitsky}
\affiliation{School of Physics and Astronomy, Raymond and Beverly Sackler
Faculty of Exact Sciences, Tel Aviv University, 69978 Tel Aviv, Israel}

\date{\today}
\begin  {abstract}
We calculate an analogue of the average phase factor of the
staggered fermion determinant at imaginary chemical potential.
Our results from the lattice agree well with the analytical predictions
in the microscopic regime for both quenched and phase-quenched QCD.  
We demonstrate that the average phase factor in the microscopic domain 
is dominated by the lowest-lying Dirac eigenvalues.  
\end{abstract}
\maketitle
 
\section{Introduction}

The numerical study of QCD at nonzero chemical potential $\mu$ is hobbled by
the complex fermion determinant, which precludes straightforward Monte Carlo
calculations based on a real measure.
Novel approaches to the problem by means of Taylor expansion
\cite{Forcrand,Allton,Gupta,Levkova}, analytic continuation
\cite{maria,owe,Az}, or reweighting \cite{Glasgow,fodor} 
are consistent in measuring the response 
to small chemical potential around the temperature of chiral restoration,
but they disagree at higher values of $\mu$. Moreover, there will be
fundamental technical difficulties at larger $\mu$ and larger volumes, 
and at smaller quark masses and lower temperatures: In all these regimes the
fluctuations in the phase of the fermion determinant will become severe
\cite{GSS,splitrev}. Understanding the behavior of the phase of the
determinant is thus of crucial importance for the interpretation of lattice
data at nonzero chemical potential (see also \cite{Ejiri}).

A common laboratory for studying features of the finite-density theory is
entered by making $\mu$ imaginary \cite{RW,AKW}.
Here the determinant is real and ordinary Monte Carlo methods can be used.
On the face of it, it appears that this would be useless for studying the
phase of the determinant since the phase is strictly zero for all
configurations. 
It is perhaps startling, then, to note that one can define and compute 
an analogue of the average phase factor at imaginary $\mu$
\cite{exp2ith-long}.  This offers an  
approach to studying the strength of the sign problem occurring at
\textit{real} chemical potential by means of simulations at imaginary
$\mu$. The advantages of this method are that one does not have to deal with  
the sign problem in order to measure its strength, and that 
one does not have to deal with eigenvalues that have spread out into the
complex plane.

The phase factor of the fermion determinant at real chemical potential is
given by  
\be
e^{2i\theta}&=&\frac{\det(D +\mu\gamma_0 +m)}{\det(D+\mu\gamma_0 +m)^*} \\
\label{phasedef}
&=&\frac{\det(D +\mu\gamma_0 +m)}{\det(D-\mu\gamma_0 +m)}.\nn
\ee
Here we have used the fact that when $\mu$ is real, the complex conjugate 
determinant is obtained by flipping its sign.  
The average phase factor at real chemical potential is then
\be
\left \langle e^{2i\theta(\mu)} \right \rangle
= \left \langle
\frac{\det(D+\mu\gamma_0+m)}{\det(D-\mu\gamma_0+m)} \right \rangle
.
\label{exp2ithmu}
\ee
As in \cite{exp2ith-long} we define the average phase factor at imaginary
$\mu$ by simply substituting $i\mu$ for $\mu$
\be
\left \langle e^{2i\theta(i\mu)}\right \rangle
\equiv \left \langle
\frac{\det(D+i\mu\gamma_0+m)}{\det(D-i\mu\gamma_0+m)}\right \rangle
,
\label{exp2ithimu}
\ee
where the parameter $\mu$ is still real. Both determinants in
\Eq{exp2ithimu} are now real. 
We present in this note a numerical study of the phase factor defined
in this manner at imaginary $\mu$. In particular we study the dependence
on the chemical potental, the quark mass, and the volume. 
We work in the microscopic domain of QCD (also called the
$\epsilon$-regime) \cite{GLeps,LS},
where the quark mass $m$ and the chemical potential are chosen to fulfill 
\be
|\mu|^2F_\pi^2 \ll \frac{1}{\sqrt{V}} \quad\textrm{and}\quad
m\Sigma        \ll \frac{1}{\sqrt{V}},
\label{mmicro}
\ee
while the Euclidean volume $V=L^3/T$ (where $T$ is the temperature) is taken
large, 
\be
V\Lambda_{\rm QCD}^4 \gg 1.
\label{mmicro2}
\ee
Here $\Sigma=\langle\bar\psi\psi\rangle$ is the chiral condensate and $F_\pi$
is the pion decay constant.%
\footnote{In fact we take $|\mu|^2F_\pi^2\sim 1/V=T/L^3$, which implies that
  $\mu/T \ll 1$ when $L$ is large. Thus the periodicity in the
  imaginary chemical potential noted by Roberge and Weiss
  \cite{RW} will never appear in the microscopic domain.} 
Formulas for the average
phase factor have recently been derived in this
regime~\cite{exp2ith-letter,exp2ith-long}.

The values of $\Sigma$ and $F_\pi$ on the lattices we use were calculated
from a two-point spectral correlation function in Refs.~\cite{DHSS}
and~\cite{DHSST-dyn-dat}.  
Given these values, we can make parameter-free comparisons of the numerical
measurements of the average phase factor with the analytical predictions. In
all cases we study the agreement is within the statistical errors. 
This confirmation of the analytic predictions at imaginary $\mu$ gives support
to the analytic results derived for real $\mu$
where direct lattice tests are harder to obtain.
Our results also show that the average phase factor in the microscopic domain
is dominated by the lowest-lying eigenvalues of the Dirac operator.

Below we will recall the theoretical predictions in the
microscopic domain and their analytic properties as functions of 
the chemical potential. Then we will compare these to the lattice data.

\section{Analytical formulas}

Our numerical results are based on data obtained in a quenched ensemble and
in an ensemble with dynamical fermions given an imaginary isospin chemical
potential. 
Here we present the microscopic formulas obtained for these two cases.

Below we work with standard unimproved staggered fermions at a coupling 
where the chiral symmetry breaking pattern is identical to that of the 
continuum theory in the sector of zero topological charge (see e.g. 
\cite{DHNR}). We therefore present the analytical predictions in the
trivial topological sector.

\subsection{Quenched theory}

\begin{figure}[ht!]
  \unitlength1.0cm
    \epsfig{file=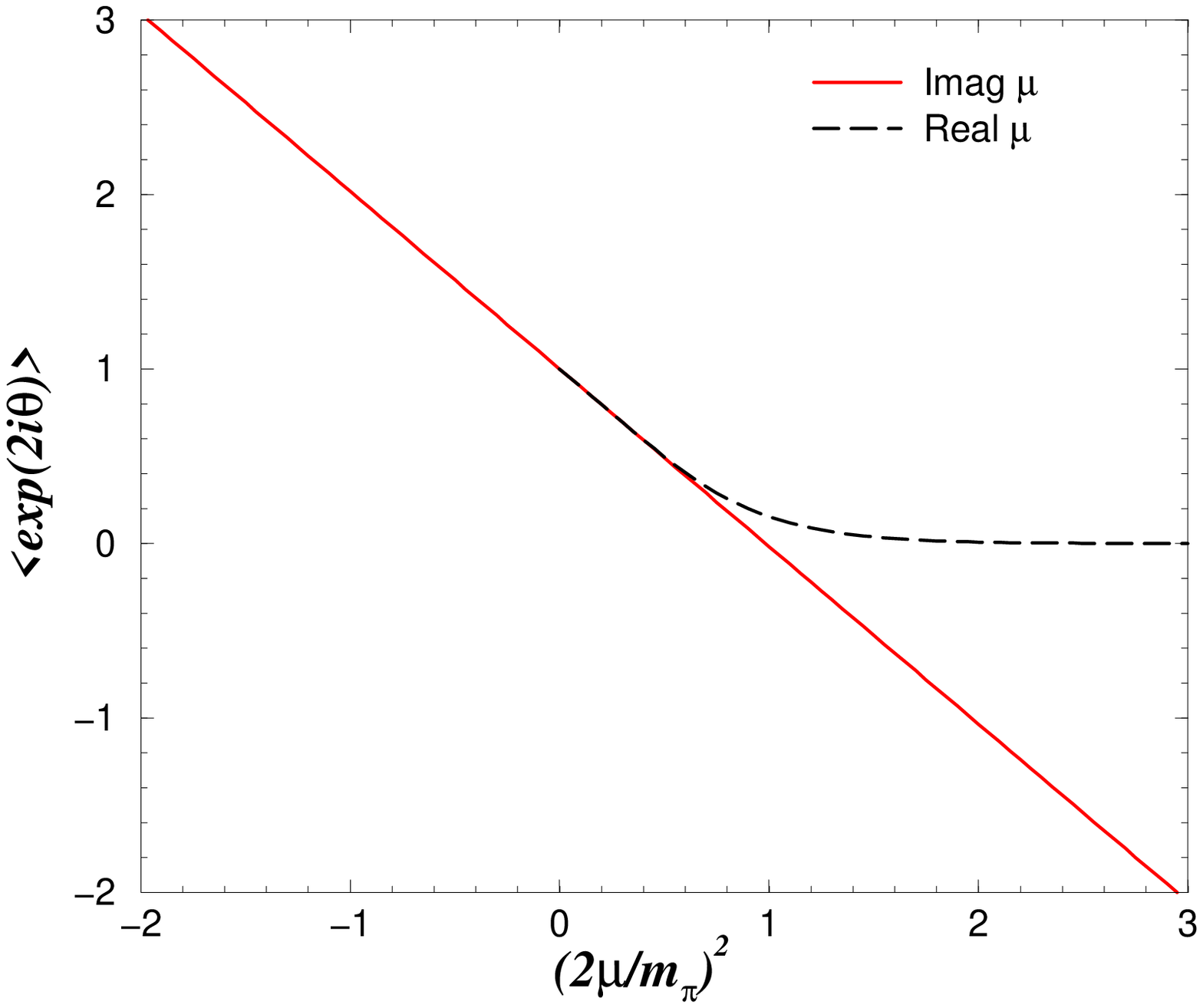,clip=,width=\columnwidth}
  \caption{ 
  \label{fig:muSQ}  
 The quenched average phase factor as a function of $(2\mu/m_\pi)^2$ for 
 $\hat{m}=3$. 
 The average phase factor from $\mu^2<0$, \Eq{phaseQ-imu}, gives the solid
 line when continued via $i\mu\to\mu$.
 This line agrees with the direct result for real $\mu$, \Eq{phaseQ-mu}, 
 when $\mu\alt m_\pi/2$.
 For larger values of $\mu$ the essential singularity in \Eq{phaseQ-mu}
 at $\mu=0$ becomes important and the analytic continuation fails to 
 reproduce the true result. The same occurs in unquenched QCD.} 
\end{figure}

The average phase factor at real chemical potential in the quenched theory is
given by \cite{exp2ith-letter,exp2ith-long}
\begin{widetext}
\be
\left\langle e^{2i\theta(\mu)} \right\rangle_{N_f=0} 
= 1-4\hatmu^2I_0(\hat{m})K_0(\hat{m})-\frac{e^{-2\hatmu^2}}{4\hatmu^2} e^{-\frac{\hat{m}^2}{8\hatmu^2}}
\int_{\hat{m}}^\infty dx\, x e^{ -\frac {x^2}{8\hatmu^2}}K_0\left ( \frac{x\hat{m}}{4\hatmu^2}\right ) \left[I_0(x)\hat{m}
I_1(\hat{m})-x I_1(x)I_0(\hat{m})\right],
\label{phaseQ-mu}
\ee
\end{widetext}
where we have defined microscopic variables via
\be
\hat m\equiv m\Sigma V\quad\textrm{and}\quad\hatmu\equiv\mu F_\pi
\sqrt{V}.
\label{scalingv}
\ee
The formula (\ref{phaseQ-mu}) cannot be continued to imaginary $\mu$ because
of the essential singularity of the last term at $\mu=0$.
The non-analyticity has its origin in the inverse 
determinant of the non-hermitian Dirac operator in
\Eq{exp2ithmu}---it is due to the eigenvalues with real part larger than the
quark mass \cite{SVbos,exp2ith-long}. For purely imaginary $\mu$ the
eigenvalues remain on the imaginary axis and are thus always inside the quark
mass. The nonanalytic term therefore is not expected to appear at imaginary
$\mu$.   
Indeed, a direct calculation at imaginary $\mu$ gives \cite{exp2ith-long}  
\be
\left\langle e^{2i\theta(i\mu)} \right\rangle_{N_f=0} 
&=& 1+4\hatmu^2I_0(\hat{m})K_0(\hat{m}) .
\label{phaseQ-imu}
\ee
As expected the analytic continuation of this result from imaginary values of
the chemical potential back to real values gives just the first two terms in
\Eq{phaseQ-mu}. For real $\mu$ the eigenvalue density outside the quark mass
is highly suppressed if $\mu\alt m_\pi/2$. 
Consequently the first two terms in (\ref{phaseQ-mu}) are dominant when $\mu
\alt m_\pi/2$ (see Fig.~\ref{fig:muSQ}).  
The measurement of the average phase factor at imaginary values of $\mu$ is 
therefore indicative of the strength of the sign problem for real $\mu \alt m_\pi/2$.

\subsection{Two dynamical flavors with isospin chemical potential} 

In this theory, also known as the phase-quenched theory, there are two
flavors of fermion with {\em opposite\/} values of chemical potential
$\pm\mu$.  The average phase of the single-flavor determinant is given (for
real $\mu$) by 
\be
\left\langle e^{2i\theta(\mu)}\right\rangle_{1+1^*} &=&\label{exp2ithPQ}
\\[2pt]
&&\hspace{-2cm}\frac{ \left\langle 
\frac{\displaystyle\det(D+\mu \gamma_0+m )}{\displaystyle\det(D+\mu \gamma_0+m)^*}\,
|\det(D+\mu \gamma_0+m )|^2\right\rangle}
{\langle |\det(D+\mu \gamma_0 +m)|^2\rangle}\, ,
\nonumber
\ee
where the averages on the right-hand side are quenched averages.
This simplifies to
\be
\left\langle e^{2i\theta(\mu)}\right\rangle_{1+1^*} &=&  
\frac{ \left\langle 
\det(D+\mu \gamma_0+m )^2\right\rangle}
{\langle |\det(D+\mu \gamma_0 +m)|^2\rangle} 
=\frac {Z_{N_f=2}}{Z_{1+1^*}}\,, \nn
\ee
a ratio of partition functions in ensembles with dynamical fermions.
These can be evaluated exactly in the microscopic regime 
\cite{LS,SplitVerb2} to obtain
\be
\left\langle e^{2i\theta(\mu)}\right\rangle_{1+1^*} 
= \frac{I_0^2(\hat{m})-I_1^2(\hat{m})}
{2\,e^{2\hatmu^2}\int_0^1 dt\, t e^{-2\hatmu^2 t^2} I_0^2(\hat m t)}
\label{pq-phase-re}
\ee
for real $\mu$, with scaling variables as defined in \Eq{scalingv} (note that
the numerator, which is $Z_{N_f=2}$, does not depend on $\mu$). In this case 
the average phase factor is analytic at $\mu=0$, because the inverse
determinant in the numerator of \Eq{exp2ithPQ} has been canceled. The average
phase factor therefore is  
\be
\left\langle e^{2i\theta(i\mu)}\right\rangle_{1+1^*} 
= \frac{I_0^2(\hat{m})-I_1^2(\hat{m})}
{2\,e^{-2\hatmu^2}\int_0^1 dt\, t e^{2\hatmu^2 t^2} I_0^2(\hat m t)}
\label{pq-phase-im}
\ee
for imaginary chemical potential $i\mu$.


\section{Numerical data}

In the lattice theory, we work with standard, unimproved staggered fermions
and introduce the chemical potential using the Hasenfratz--Karsch
prescription \cite{HK}. 
As in the continuum, each operator $D\pm i\mu\gamma_0$ is
anti-hermitian,
and anticommutes with $\gamma_5$. Each operator's eigenvalues
therefore come in pairs of opposite sign on the imaginary axis.
For each gauge field configuration we measure the two sets of
eigenvalues defined by 
\be
[D(A)+i\mu\gamma_0]\psi_{+}^{(j)}=i\lambda_{+}^{(j)}\psi_{+}^{(j)},\\[3pt]
[D(A)-i\mu\gamma_0]\psi_{-}^{(j)}
= i\lambda_{-}^{(j)}\psi_{-}^{(j)}.
\ee
Thus $\det(D\pm i\mu\gamma_0+m)=\prod_j(i\lambda_{\pm}^{(j)}+m)$.
Combining positive with negative $\lambda$'s, the average phase factor is given by
\be
\left\langle e^{2i\theta(i\mu)}\right\rangle
&=& \left\langle \prod_{j=1}^{\infty}
\frac{\lambda_+^{(j)\,2}+m^2}{\lambda_-^{(j)\,2}+m^2}\right\rangle
\equiv\langle\Phi_{+-}\rangle
\label{infprod1}\\[3pt]
&=& \left\langle \prod_{j=1}^{\infty}
\frac{\lambda_-^{(j)\,2}+m^2}{\lambda_+^{(j)\,2}+m^2}\right\rangle
\equiv\langle\Phi_{-+}\rangle
\label{infprod2}
\ee   
where the products run over positive $\lambda$'s only.  The equality of
$\langle\Phi_{+-}\rangle$ and~$\langle\Phi_{-+}\rangle$ follows from charge
conjugation symmetry, which exchanges $\lambda_+$ with $\lambda_-$ and is
true in both the quenched and the phase-quenched theory (but not in
the unquenched theory). With finite statistics, the two estimators
$\Phi_{+-}$ and~$\Phi_{-+}$ give distinct measurements of the average phase.

\subsection{Quenched theory}

Our quenched eigenvalue data are a combination of data used in
Ref.~\cite{DHSS} with new data.
We simulated the SU(3) gauge theory with the plaquette action at $\beta=5.7$
on lattices with $8^4$ and $12^4$ sites.  For each gauge configuration we 
calculated the smallest 24 pairs of positive eigenvalues
$\lambda_\pm^{(j)}$.

\begin{figure}[ht]
  \unitlength1.0cm
    \epsfig{file=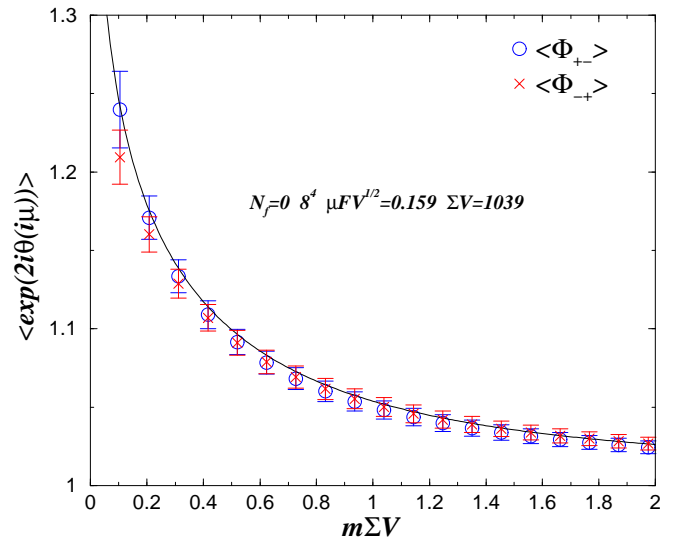,clip=,width=\columnwidth}
  \caption{ 
  \label{fig:Q8i4}  
 The average phase factor calculated via \Eq{infprod1} (circles) and via
 \Eq{infprod2} 
 (crosses) from the lowest 24 positive eigenvalues of the staggered
 Dirac operator on an $8^4$ lattice, for fixed chemical potential $a\mu=0.01$ as a
 function of the scaled quark mass; 4000 configurations. The full line is the
 analytical microscopic prediction using the parameters determined in
 Ref.~\cite{DHSS}.} 
\end{figure}

On the smaller lattice, we chose two values of the chemical potential: 
$a\mu=0.01$ and $a\mu=0.1$.
In Ref.~\cite{DHSS} we determined $F_\pi$ and $\Sigma$,
giving the scaled values $\hatmu=0.159$ and
$\hatmu=1.59$, respectively; also $\Sigma V/a=1039$, which permits converting
the lattice masses $am$ into scaled masses $\hat m$. 
With these values we can make a parameter-free comparison between the
analytical prediction~(\ref{phaseQ-imu}) and the 
measurement of the (truncated) average phase factor.
The result for $a\mu=0.01$, based on data from Ref.~\cite{DHSS}, is shown in
Fig.~\ref{fig:Q8i4}. 
The agreement is well within the statistical error bars.
  
\begin{figure}[ht]
  \unitlength1.0cm
    \epsfig{file=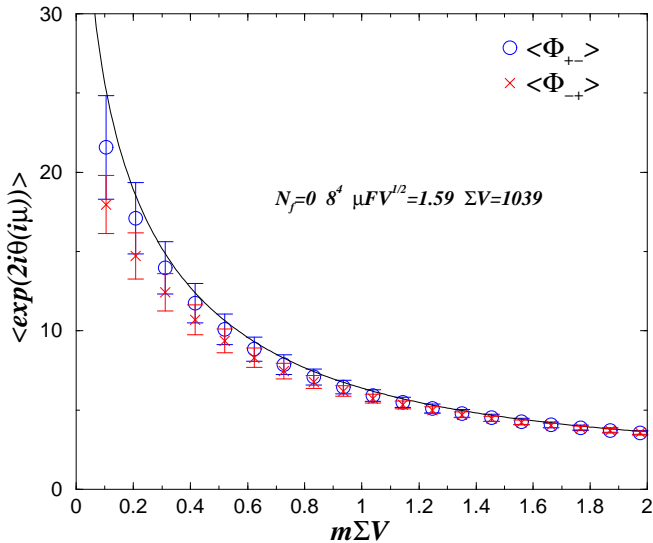,clip=,width=\columnwidth}
  \caption{ 
  \label{fig:Q8i4mu0p1}  
 As in Fig.~\ref{fig:Q8i4} but with $a\mu=0.1$; 6068 configurations.}
\end{figure}

The value $a\mu=0.01$ 
was chosen in Ref.~\cite{DHSS} to ensure that the scaling variable $\hatmu$ was
roughly 1/10, since this facilitates the determination of $F_\pi$. At such
small values of $\mu$ the shift of the eigenvalues due to the
imaginary chemical potential is small compared to the average level spacing.%
\footnote{For a real $\mu$ of this order the width of the strip of eigenvalues in the complex plane is much smaller than the average level spacing along the imaginary axis \cite{SplitVerb2,O,AOSV}.}
This was our motivation in running a new quenched 
simulation in order to calculate eigenvalues with the larger value $a\mu=0.1$. The result is shown in  Fig.~\ref{fig:Q8i4mu0p1}; again the agreement with \Eq{phaseQ-imu} is good. Note that in this case $\mu L=0.8$ which is pushing the limit of the microscopic domain.   

\begin{figure}[ht]
  \unitlength1.0cm
    \epsfig{file=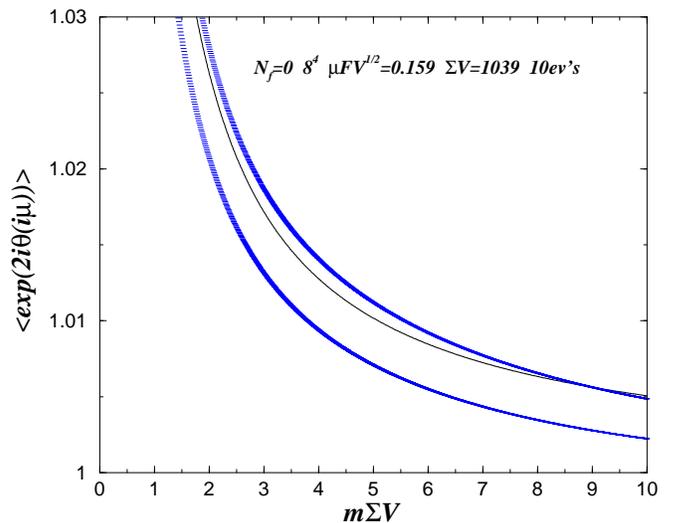,clip=,width=\columnwidth}
  \caption{ 
  \label{fig:Largem}  
Same eigenvalue data as in 
Fig.~\ref{fig:Q8i4}, but retaining only 10 eigenvalues.
As the quark mass approaches the largest retained eigenvalue
($\lambda^{(10)}\Sigma V\sim 30$) 
we observe that the error band drops
below the analytic curve (thin line). }  
\end{figure}

Since $m$ is not a dynamical parameter of the quenched simulation, the data points
for different values of $m$ are highly correlated, and the variation among
them is systematic, not statistical.  In effect, $m$ determines which
eigenvalues contribute the most to the average phase in Eqs.~(\ref{infprod1})
and~(\ref{infprod2}). When $m$ is small, all eigenvalues contribute to the
phase; the smallest eigenvalues cause large fluctuations, while the large
eigenvalues show but little dependence on $\mu$ and thus 
contribute little to the ratio. When $m$ is large it eliminates the 
effect of small eigenvalues on the phase.  
Therefore as $m$ approaches the largest of the calculated eigenvalues 
there is a systematic deviation from the analytic curve.
We illustrate this by cutting the number of calculated eigenvalues
to 10. Figure \ref{fig:Largem} shows the deviation from the
analytic curve as $m\Sigma V $ approaches $\lambda^{(10)}\Sigma V\sim 30$.

\begin{figure}[ht]
  \unitlength1.0cm
    \epsfig{file=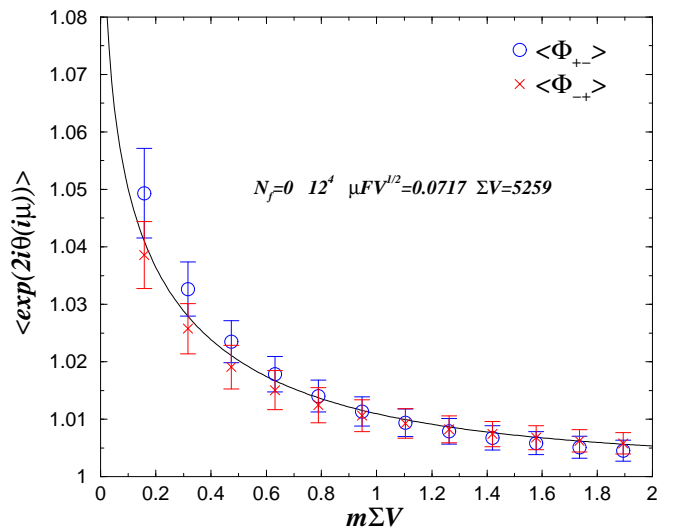,clip=,width=\columnwidth}
  \caption{ 
  \label{fig:Q12i4}  
 As in Fig.~\ref{fig:Q8i4}, but for the $12^4$ lattice; 2000 configurations.}
\end{figure}

For the larger lattice, with $12^4$ sites, we use data from Ref.~\cite{DHSS} with $a\mu=0.002$.
In Ref.~\cite{DHSS} we
determined that $\hatmu=\mu F_\pi\sqrt{V}=0.159$ and $\Sigma V/a=5259$. Again we can
compare the analytical prediction (\ref{phaseQ-imu}) to the  
measurement of the truncated average phase factor without free
parameters, as shown in Fig.~\ref{fig:Q12i4}.  The agreement is clear,
although statistically weaker than for the smaller lattice. 

\subsection{Two dynamical flavors}

For the two-flavor theory with imaginary isospin chemical potential 
we use eigenvalue data calculated in the long runs of
Ref.~\cite{DHSST-dyn-dat}, which should be consulted for numerical details.
The lattice theory is based on the plaquette
action and unimproved staggered fermions, simulated at $\beta=4.2$ with
volume $6^4$.  We have data for two masses, $am=0.002$ and~0.005,
and the imaginary chemical potential coupled to isospin is
$a\mu=0.0055$. 
A determination \cite{DHSST-dyn-dat} of $aF_\pi$ gives
$\hatmu=0.1338$.

For the smaller quark mass, $am=0.002$, we calculated \cite{DHSST-dyn-dat}
that $\hat m=m\Sigma V=3.318$. From the lowest 24 eigenvalues we measure 
\be
\left\langle 
\Phi_{+-}\right\rangle_{1+1^*} 
& = & 
1.0092\pm0.0020\\
\left\langle 
\Phi_{-+}\right\rangle_{1+1^*} & = & 
1.0082\pm0.0020.
\ee   
The analytical prediction for the average phase factor (\ref{pq-phase-im}) 
is
\be
\left\langle e^{2i\theta(i\mu)}\right\rangle_{1+1^*} = 1.0094,
\ee
in agreement with the numerical averages.

For the larger quark mass ($ma=0.005$) we have $m\Sigma V=8.295$. 
From the lowest 24 eigenvalues we compute  
\be
\left\langle 
\Phi_{+-}\right\rangle_{1+1^*} & = & 1.0036\pm0.0030\\
\left\langle 
\Phi_{-+}\right\rangle_{1+1^*} & = & 1.0032\pm0.0030.
\ee
The analytical formula (\ref{pq-phase-im}) gives
\be
\left\langle e^{2i\theta(i\mu)}\right\rangle_{1+1^*} = 1.0041.
\ee
Here also we see satisfactory agreement with theory, although the
deviation from unity is not really significant.

\section{Conclusions}

We have computed an analogue of the complex phase factor of the
fermion determinant at imaginary values of the chemical potential. The 
results have been used as a parameter-free test of the predictions from 
the microscopic domain of QCD. In the quenched as well as in the dynamical 
cases studied the numerical and analytical results are in agreement. 
Furthermore, we have demonstrated that the phase factor in the microscopic
domain is dominated by the low-lying Dirac eigenvalues.

We hope that this first numerical study of the phase factor at imaginary
chemical potential will encourage studies also outside the microscopic
domain.

\begin{acknowledgments}
We wish to thank P.~H.~Damgaard, P.~de~Forcrand, and
J.~J.~M.~Verbaarschot for discussions. This work was
supported by the Israel Science Foundation under grant no.~173/05 (BS) and by
the Carlsberg Foundation (KS). BS thanks the Niels Bohr
Institute for its hospitality.
\end{acknowledgments}

\end{document}